\documentclass[twocolumn,english,conference]{IEEEtran}
\usepackage[T1]{fontenc}
\usepackage{babel}
\usepackage{graphicx}

\makeatletter

\usepackage{algorithm} 
\usepackage{algpseudocode}
\usepackage{float}
\usepackage{amsmath}
\makeatother

\begin{document}

\title{Social Network Generation and Role Determination Based on Smartphone Data}

\author{\IEEEauthorblockN{Mustafa \.{I}lhan Akba\c{s}, Matthias R. Brust\\Advisor: Damla Turgut\\
  \IEEEauthorblockA{Department of Electrical Engineering and Computer Science\\ University of Central Florida, Orlando, FL\\ Email:
\{miakbas,mbrust,turgut\}@eecs.ucf.edu}}}
\maketitle

\begin{abstract}
We deal with the problem of automatically generating social networks
by analyzing and assessing smartphone usage and interaction data. We start by assigning weights to the different types of interactions such as messaging, email, phone calls, chat and physical proximity. Next, we propose a ranking algorithm which recognizes the pattern of interaction taking into account the changes in the collected data over time. Both algorithms are based on recent findings from social network research.
\end{abstract}

\section{Introduction}

\IEEEPARstart{O}{ver} a billion people frequently use social network
web sites. The success of these web sites and related applications
lies in the convenient connection of the user to a digital social network of friends and acquaintances. As a result of this process, the user creates a digital version of his or her real-life social network. However, the creation and maintenance of the virtual social network requires a large number of interactions, as it is the user's task to establish or remove the connections to friends and sites of interests.

While these explicit interactions give the user the impression of full
control over the virtual social network, they are time-consuming.
Moreover users are confronted with the need to organize the internal
structure of their personal virtual social network. Some social sites
such as Facebook offer a partial support for this process for family
members and spouses. Other social networks try to offer a full support
by enabling the creation of social circles as desired (Google+). In
either case, the user assigns the friends and acquaintances to an
individual structure.

In our ongoing work, we deal with the problem of automatically generating
the user's social network by using different sources of interaction data 
such as physical proximity, messaging, phone calls and video chats. The
main challenge of this work is to make inferences from a limited interaction
history. We approach this problem by first evaluating the interactions
according to their types. Then these values are used to rank friends of
users and to find the friendship levels in the social network of each user.

\section{Social Network Generation}

Smartphones allow us to trace multiple types of interactions between a user and the members of its social network. For instance, text messages, calls and email conversations are stored as history on the device or in the cloud. Moreover, smartphones are equipped with many sensors, which sense, evaluate and record even more information about the user, the environment and contacts.
For instance the proximity of two users is detected by acoustic sensors
or the location information is collected by GPS receivers. 

Usage of all available smartphone data enables us to infer a social
network. However, one of the critical problems is that the generation
of the network cannot be measured in terms of accuracy or other measures
that would allow us to state that the collected data sufficiently
describes the corresponding social network. The answers of the questions
related to network dynamics such as when a node is a node and when
a link really exists are visible influences on our model.

The proximity and mobility data collection part of the approach is
based on our the algorithms developed on our previous work \cite{Akbas11}.
We assume basic interaction information of the format {[}UserA, UserB, 
	timeStamp, interactionType{]}. Instead of plainly reading and transforming the
interaction data, these structural artifacts are used as structural
framework. Additional to the information derived from the interaction
data, this framework gives information on how the nodes are located
in the network and links have been created.

\subsection{Interaction evaluation}

The interaction data derived from mobile devices and sensors need
to be evaluated in terms of their importance. Intuitively, an email
to a person appears to be more distant than calling the same person
and a phone call is shown to be less effective for personal relationships than meeting with the other party in person\cite{Dunbar10}. In order to take into account these differences, we suggest assigning specific weights to different types of interactions. Weight assignment results
in the ability to change the weight according to experience or context
and to include additional interaction types as needed. The interaction
values are defined in our approach as follows: 

\[
i_{A,B}=\alpha\cdot F(T)+\beta\cdot V(T)+\gamma\cdot nP(T)+\delta\cdot E(S)
\]

\noindent where $P(T)$, $V(T)$ and $F(T)$ denote the number of
times respectively a phone call, a video conference and a face-to-face
interaction occurred for a particular amount of time, $T$. $E(S)$
denotes the number of e-mails or text messages with size $S$. 

Each interaction type has a different constant ($\alpha,\beta,\gamma,\delta$),
which reflects the variety in effects of different interaction types
on personal relations. Formulation of this equation and finding the
exact values of constants is one of the next steps of our work. Due
to the nature of social sciences, these values may change depending
on various factors such as the social group under investigation. Our
approach provides the means to utilize results of works in social
sciences such as the study of Okdie et al.~\cite{okdie11} on comparison
of face-to-face communication versus online communication.

\subsection{Ranking and Role Determination}

The interactions between friends correlate in number with the strength
of friendship \cite{Dunbar10}. Therefore, after determining the friends 
of a person and evaluating the interactions, our system also ranks the 
friends to find different levels of friendship in the social network 
of users.

We say that the friend with a larger interaction
value has a {\em win} against the friend with lower interaction value for that time period. Therefore the sports ranking methods, in which
teams win/loose against each other, can be utilized to rank friends.
However, a simple sports ranking method, which uses win percentage
would not satisfy the particularities of our application. Colley \cite{Colley03} and Massey \cite{Massey97} are two of the the most important sports ranking methods, which take the history and current ranking into consideration. Chartier et al.~\cite{Chartier11} made a sensitivity analysis of these methods and concluded that the Colley and Massey methods are insensitive to small changes, which is desirable for social network ranking. For instance when a person spends most of the time in a week with a new friend, this new friend wouldn't suddenly become one of the closest friends of that person.  Colley's method is based only on results from the field whereas Massey method utilizes actual game scores and homefield advantage, which have no correspondence in social networks. Therefore, we had chosen the Colley method as the basis for our ranking.

The Colley method of sports ranking can be defined by a linear system {[}{*}{]}, $C\vec{{r}}=\vec{{b}}$,
where \textbf{$\vec{{r}}_{n\times1}$} is a column-vector of all the
rating \textbf{$\vec{{r}}_{i}$}, \textbf{$\vec{{b}}_{n\times1}$}
is the right-hand-side vector defined as follows:

\[
\vec{{b}}_{i}=1+(w_{i}-l_{i})/2
\]
$C_{n\times n}$is called the Colley coefficient matrix and defined
as follows:

 \[  
	C_{ij} =   
		\begin{cases}    
			2+t_{i} & i = j \\
			-n_{ij} & i \neq j   
		\end{cases} 
 \]

The scalar \textbf{$n_{ij}$} is the number of times friends \textbf{$i$}
and \textbf{$j$} are compared to each other, \textbf{$t_{i}$} is
the total number of comparisons for friend \textbf{$i$}, \textbf{$w_{i}$}
is the number of wins and \textbf{$l_{i}$} is the number of losses
for \textbf{$i$}. It can be proved that the Colley system $C\vec{{r}}=\vec{{b}}$
always has a unique solution since $Cn\times n$ is invertible. Then
the rank is defined as follows \cite{Chartier11}:

\begin{equation} r_{i}=\frac{1+n_{w,i}}{2+n_{total,i}} \nonumber \end{equation}

In contrast to traditional methods, the initial rating of any
friend with no changes is equal to $\frac{1}{2}$, which is the median
value between 0 and 1. Depending on the comparisons, a win increases
and loss decreases the value of $r$. This approach results in a system
less sensitive to changes. 

\begin{figure}
\centering
\includegraphics[width=0.45\textwidth]{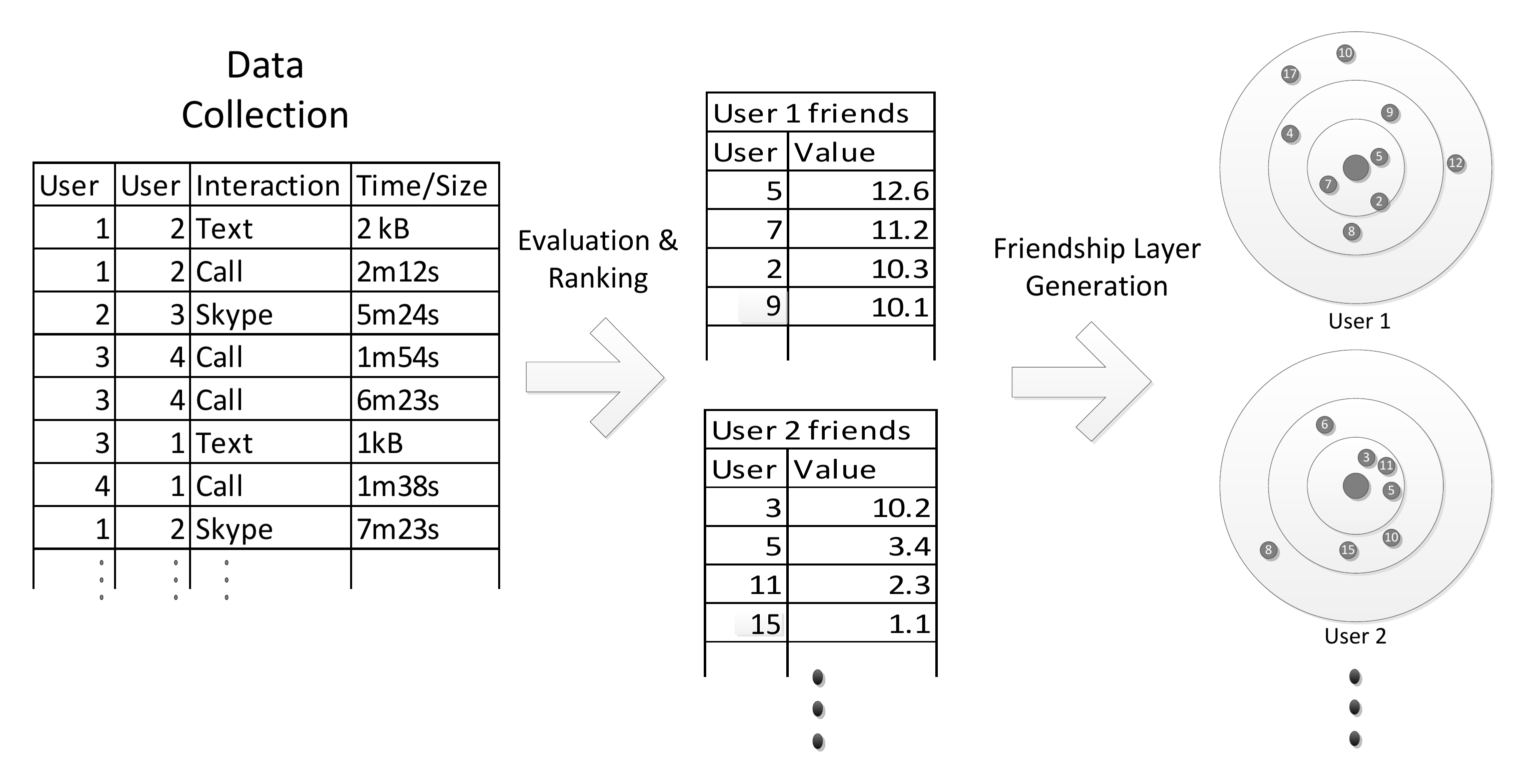}
\caption{Overview of social network generation for users.}
\label{Overview}
\end{figure}

Dunbar et al. \cite{Dunbar10} showed that the social network and
the number of friends of a person have a common pattern. The types
of social groups formed according to ties in our social networks have
clear boundaries between each other \cite{zhou2005}. Social networks
are hierarchically organized in discretely sized groups, which are
also refered as circles. The inner most circle includes up to five
people and the sizes of circles increase by a factor of three. The
discrete groups formed in our approach as friends
are ranked and the circle of a friend is decided according
to this organization as given in Fig. \ref{Overview}.

\section{Conclusion}

The approach outlined in this paper aims to infer the social networks
of individuals by assessing their smartphone data. We propose a data
collection method, an interaction evaluation function and a ranking
method in accordance with the research findings in social networks.
Our future work includes testing our approach by using real-life smartphone data.

\bibliographystyle{ieeetr}
\bibliography{IEEEfull,RankingSocNet}

\end{document}